\documentclass{PoS}

\newcommand{\comment}[1]{}
\newcommand{\lr}[1]{ \left( #1 \right) }
\newcommand{\lrs}[1]{ \left[ #1 \right] }
\newcommand{\lrc}[1]{ \left\{ #1 \right\} }
\newcommand{\vev}[1]{ \langle \, #1 \, \rangle }

\newcommand{\tr}{ {\rm Tr} \, }

\newcommand{\sign}[1]{ {\rm sign}\left( #1 \right)}
\newcommand{\expa}[1]{ \exp{\left( #1 \right)} }

\title{Metropolis updates for Diagrammatic Monte-Carlo algorithms from Schwinger-Dyson equations}

\ShortTitle{Diagrammatic Monte-Carlo from Schwinger-Dyson equations}

\author{\speaker{P.~V.~Buividovich}%
         \thanks{This work was supported by the S.~Kowalevskaja award from the Alexander von Humboldt foundation.}\\
        Institute for Theoretical Physics, Regensburg University, D-93053 Regensburg, Germany\\
        E-mail: \email{pavel.buividovich@physik.uni-regensburg.de}}

\abstract{We describe a general recipe for constructing Metropolis updates for Diagrammatic Monte-Carlo (DiagMC) algorithms, based on the Schwinger-Dyson equations in quantum field theory. This approach bypasses explicit duality transformations, enumeration or classification of diagrams and can be used for lattice quantum field theories with unknown or complicated dual representations (such as non-Abelian lattice gauge theories). DiagMC algorithms constructed in this way can still be plagued by the sign problem, which is, however, completely different from the sign problem in conventional Monte-Carlo simulations and has its origin in cancellations between diagrams with positive and negative weights. To test the presented approach, we apply DiagMC to calculate the first 7 orders of $1/N^2$ expansion in the quartic matrix model and find good agreement with analytic results, with the exception of the close vicinity of the critical coupling where the critical slowing down sets in.}

\FullConference{34th annual International Symposium on Lattice Field Theory\\
                 24-30 July 2016\\
                 University of Southampton, UK}

\begin{document}
\sloppy

\section{Introduction}
\label{sec:introduction}

 Diagrammatic Monte-Carlo algorithms have turned out to be extremely useful for the first-principle studies of quantum systems for which ordinary configuration-space Monte-Carlo simulations based on the sampling of field configurations become impossible due to the sign problem. Typically, Diagrammatic Monte-Carlo algorithms are constructed by using some explicit diagrammatic representation of strong- or weak-coupling expansion series. One constructs certain ergodic set of transformations on the diagram space and accepts each of the transformations in a Metropolis algorithm according to the ratio of the weights of the initial and transformed diagrams \cite{Prokofev:08:1, Prokofev:01:1}. Explicit forms of perturbative expansions are typically easy to construct for simple condensed matter systems with on-site or static inter-electron interactions \cite{Prokofev:08:1, Prokofev:01:1} or for Abelian gauge theories \cite{Gattringer:14:1}. For non-Abelian lattice gauge theories or principal chiral models, however, both strong- and weak-coupling expansions become quite complicated at high orders, and it is difficult to classify and parameterize all the admissible diagrams and to calculate their weights (see \cite{Gattringer:16:1} for some recent work in this direction).

 On the other hand, it is known that weak- or strong-coupling expansions in quantum field theories can be obtained by iteratively solving the Schwinger-Dyson equations. In contrast to explicit dual representations, Schwinger-Dyson equations are easy to derive in any QFT with continuous field variables. E.g. for non-Abelian gauge theories these are the Migdal-Makeenko loop equations \cite{Migdal:81:1}. In these Proceedings we demonstrate how a suitable set of Metropolis updates for a DiagMC algorithm can be obtained directly from Schwinger-Dyson equations. After presenting the general prescription in Section \ref{sec:stochastic_linear}, in Section \ref{sec:implementation} we discuss the practical implementations of the algorithm. Finally, in Section \ref{sec:phi4sd_nonplanar} we illustrate the presented approach on the example of the Schwinger-Dyson equations in the quartic Hermitian matrix model, and test it by calculating the first 7 orders of $1/N^2$ expansion of the free energy. In contrast to the stochastic solution of Schwinger-Dyson equations presented by the author in \cite{Buividovich:10:2,Buividovich:11:1}, the algorithm presented here works even very close to the critical coupling at which the expansions being sampled diverge (of course, exhibiting the expected critical slow-down). Application of this algorithm to principal chiral models and non-Abelian lattice gauge theories will be described in detail in forthcoming publications.

\section{Stochastic solution of linear equations}
\label{sec:stochastic_linear}

 Schwinger-Dyson equations can always be represented as infinite-dimensional linear equations on the space of field correlators $\phi\lr{X} = \vev{\phi\lr{x_1} \ldots \phi\lr{x_n}}$ (which include both connected and disconnected contributions) with all possible values of $X = \lrc{x_1, \ldots, x_n}$ and $n$:
\begin{eqnarray}
\label{sd_linear_general}
 \phi\lr{X} = b\lr{X} + \sum\limits_Y A\lr{X | Y} \phi\lr{Y},
\end{eqnarray}
where $A\lr{X | Y}$ is some theory-dependent linear operator and $b\lr{X}$ represents the ``source'' terms in the Schwinger-Dyson equations (typically, the contact delta-function term in the lowest-order Schwinger-Dyson equation). In Section~\ref{sec:phi4sd_nonplanar} we give a particular example of such representation of Schwinger-Dyson equations for the quartic matrix model.

 The solution $\phi\lr{X}$ of (\ref{sd_linear_general}) can be written as formal geometric series in $A$:
\begin{eqnarray}
\label{sd_linear_powerseries_explicit}
 \phi\lr{X} = \sum\limits_{n=0}^{+\infty} \sum\limits_{X_0} \ldots \sum\limits_{X_n} \delta\lr{X, X_n} A\lr{X_n | X_{n-1}} \ldots A\lr{X_1 | X_0} b\lr{X_0} .
\end{eqnarray}
In case one truncates the strong- or weak-coupling expansion at some finite order, the above series contain only a finite number of terms, which correspond to a finite number of diagrams contributing up to given expansion order.

 We propose to evaluate the series (\ref{sd_linear_powerseries_explicit}) by stochastically sampling the sequences of variables $\mathcal{S} = \lrc{X_n, \ldots, X_0}$ with arbitrary $n$ in (\ref{sd_linear_powerseries_explicit}) with probability
\begin{eqnarray}
\label{sd_linear_sampling_prob}
 w\lr{\mathcal{S}}
 =
 \mathcal{N}_w^{-1} |A\lr{X_n | X_{n-1}}| \ldots |A\lr{X_1 | X_0}| |b\lr{X_0}| .
\end{eqnarray}
In order to simplify the notation in what follows, let us also define the quantities
\begin{eqnarray}
\label{n_def}
 \mathcal{N}\lr{Y} = \sum\limits_X |A\lr{X | Y}|, \quad \mathcal{N}_b = \sum\limits_X |b\lr{X}| .
\end{eqnarray}

 The solution $\phi\lr{X}$ to the system (\ref{sd_linear_general}) can be obtained as a histogram of the last variable $X_n$ in the sequence $\lrc{X_n, \ldots, X_0}$, where each occurrence of $X_n$ is weighted with the sign $\sigma\lr{X_n, \ldots, X_0} = \sign{A\lr{X_n | X_{n-1}}} \ldots \sign{A\lr{X_1 | X_0}} \sign{b\lr{X_0}}$. This sign reweighting puts certain limitations on the use of the method, since we are effectively sampling the series in which the linear operator $A\lr{X | Y}$ is replaced by $|A\lr{X | Y}|$ and which typically have a smaller radius of convergence. In the worst case, the expectation value of the reweighting sign $\sigma\lr{X_n, \ldots, X_0}$ can decrease exponentially with $n$, thus limiting the maximal expansion order which can be sampled by the algorithm.

 In order to sample the sequences $\mathcal{S}$ with probability (\ref{sd_linear_sampling_prob}), we use the Metropolis-Hastings algorithm with the following updates:
\begin{description}
\label{desc:transitions}
 \item[Add element:] With probability $p_+$ add a new element $X_{n+1}$ to the sequence $\lrc{X_n, \ldots, X_0}$, where the probability distribution of $X_{n+1}$ is $\pi\lr{X_{n+1} | X_n} = |A\lr{X_{n+1} | X_n}|/\mathcal{N}\lr{X_n}$. Multiply the sign variable $\sigma$ by $\sign{A\lr{X_{n+1} | X_n}}$.
 \item[Remove element:] If the sequence contains more than one element, with probability $1 - p_+$ remove the last element $X_n$, thus transforming the sequence $\lrc{X_n, X_{n-1}, \ldots, X_0}$ into $\lrc{X_{n-1}, \ldots, X_0}$. Multiply the sign variable $\sigma$ by $\sign{A\lr{X_n | X_{n-1}}}$.
 \item[Restart:] If the sequence contains only one element $X_0$, replace it by $X_0'$ with probability $1 - p_+$. The probability distribution of $X_0'$ is $\pi\lr{X_0'} = |b\lr{X_0'}|/\mathcal{N}_b$. Set the sign variable $\sigma$ to $\sign{b\lr{X_0'}}$.
 \end{description}
Since for the above updates the detailed balance condition $\pi\lr{\mathcal{S}' \rightarrow \mathcal{S}} = \pi\lr{\mathcal{S} \rightarrow \mathcal{S}'}$ for transition probabilities between the sequences $\mathcal{S}$ and $\mathcal{S}'$ is in general not satisfied, they should be then accepted or rejected with the probability $\alpha\lr{\mathcal{S} \rightarrow \mathcal{S}'} = \min\lr{1, \frac{w\lr{\mathcal{S}'} \pi\lr{\mathcal{S' \rightarrow S}}}{w\lr{ \mathcal{S}} \pi\lr{\mathcal{S  \rightarrow S'}}}}$ \cite{Hastings:70:1}. For the three updates defined above we find the following acceptance probabilities:
\begin{eqnarray}
\label{acceptance_probabilities}
 \alpha_{add}     =  \frac{\mathcal{N}\lr{X_n} \lr{1 - p_+}}{p_+} , \quad
 \alpha_{remove}  =  \frac{p_+}{\mathcal{N}\lr{X_{n-1}} \lr{1 - p_{+}} }, \quad
 \alpha_{restart} = 1 .
\end{eqnarray}
After some algebra one can express the overall acceptance rate as $\alpha = 2 \vev{\min\lr{ p_+, \lr{1 - p_+} \mathcal{N}\lr{X_n} }} + \lr{1 - p_+} \frac{\mathcal{N}_b}{\mathcal{N}_w}$.  In order to reach the optimal performance of the algorithm with acceptance close to unity, one can start the simulations with some initial value of $p_+$, calculate the expectation values $\vev{\min\lr{ p_+', \lr{1 - p_+'} \mathcal{N}\lr{X_n} }}$ for several trial values $p_+'$ and then select the value which maximizes the acceptance.

 It is also useful to calculate the normalization factor $\mathcal{N}_w$ which relates the (sign weighted) histogram of the last element $X_n$ in the sequences $\mathcal{S} = \lrc{X_n, \ldots, X_0}$ and the actual solution $\phi\lr{X}$ of the system (\ref{sd_linear_general}). From (\ref{sd_linear_sampling_prob}) and (\ref{sd_linear_general}) one can obtain a linear equation for $\mathcal{N}_w$: $\mathcal{N}_w = \mathcal{N}_w \sum\limits_X w\lr{X} = \mathcal{N}_w \sum\limits_{X,Y} |A\lr{X | Y}| w\lr{Y} + \sum\limits_X |b\lr{X}| = \mathcal{N}_w \sum\limits_Y \mathcal{N}\lr{Y} w\lr{Y} + \mathcal{N}_b$, where $w\lr{X}$ is the probability distribution of the last element $X_n$ in the sequences $\mathcal{S}$, with any sign $\sigma$. From the above equation we can finally express $\mathcal{N}_w$ as
\begin{eqnarray}
\label{norm_factor_linear}
 \mathcal{N}_w = \frac{\mathcal{N}_b}{1 - \vev{\mathcal{N}\lr{X_n}}} .
\end{eqnarray}

\section{Practical implementation of the stochastic solution of Schwinger-Dyson equations}
\label{sec:implementation}

 Typically, for Schwinger-Dyson equations of the general form (\ref{sd_linear_general}) the coefficients $b\lr{X}$ and $A\lr{X | Y}$ are different from zero for only a few values of $X$ and $Y$. Moreover, the space of $X$ variables is in most cases infinite. Therefore it is not possible but also not necessary to store $b$ and $A$ entirely in computer memory. Rather, it is convenient to implement them as a user defined functions which return only nonzero values of $b\lr{X_0}$ and $A\lr{X_{n+1} | X_n}$ for a given $X_n$. Given all nonzero values of $A\lr{X_{n+1} | X_n}$, the algorithm should calculate $\mathcal{N}\lr{X_n}$ and the acceptance probability $\alpha_{add}$ for the \textbf{Add index} transition. The calculated values $\mathcal{N}\lr{X_n}$ can be stored as an ordered sequence $\lrc{\mathcal{N}\lr{X_n}, \mathcal{N}\lr{X_{n-1}}, \ldots, \mathcal{N}\lr{X_0}}$ (which, in fact, has the stack structure). Since at each step of the random process either a new element $X_{n+1}$ is attached to the sequence or the last element $X_n$ is removed, and all other elements are not changed, one can re-use the previously calculated values of $\mathcal{N}\lr{X_{n-1}}$ for the calculation of the acceptance probability $\alpha_{remove}$ of the \textbf{Remove index} transition.

 Let us now turn to the optimal practical way of saving the sequences $\mathcal{S} = \lrc{X_n, \ldots, X_0}$. In practice, the variables $X$ are themselves the sequences of some variables (e.g. lattice coordinates or momenta), and saving them entirely in computer memory is quite impractical. Let us note, however, that for a given sequence the next proposed element $X_{n+1}$ only depends on $X_n$. Moreover, as a consequence of the sparseness of the coefficients $b\lr{X}$ and $A\lr{X | Y}$, only a rather small set of new elements $X_{n+1}$ might be proposed for a given $X_n$. Thus in practice it is possible to enumerate all possible updates for a given set of Schwinger-Dyson equations and characterize them by some small set of parameters (in most cases, only a few integer or real numbers). Instead of keeping the whole sequence $\lrc{X_n, \ldots, X_0}$ in memory, one can store only the topmost element $X_n$ as well as the history of updates $X_0 \rightarrow X_1, \ldots, X_{n-1} \rightarrow X_n$ which led to a given sequence. If the \textbf{Remove index} is proposed and accepted by the algorithm, the last update should be ``undone''.

 Sometimes it can be also useful to graphically interpret different terms in the Schwinger-Dyson equations as some transformations on the external legs of the diagrams being sampled. The Monte-Carlo process described above can be then thought of as a graphical editor for diagram drawing which has several commands for updating the diagrams, such as drawing a line or a vertex. At each Monte-Carlo step one either chooses a random update command with some probability, or undoes the previous update. It is clear then that the state of the algorithm can be completely described by the current diagram and the history of commands which were used to draw it.

\section{Test case: high orders of $1/N$ expansion for the $\phi^4$ matrix model}
\label{sec:phi4sd_nonplanar}

 In this Section we test the DiagMC algorithm described above on the simplest example of the quartic Hermitian matrix model, which is specified by the partition function
\begin{eqnarray}
\label{phi4_matrix_model}
 \mathcal{Z}\lr{\lambda}
 =
 \int \mathcal{D} \phi \expa{-\frac{N}{2} \tr\phi^2 + \frac{\lambda N}{4} \tr\phi^4 } ,
\end{eqnarray}
where we integrate over $N \times N$ Hermitian matrices $\phi$. A full set of linear Schwinger-Dyson equations of the form (\ref{sd_linear_general}) for this model can be written in terms of the multi-trace expectation values expanded into power series in $1/N^2$:
\begin{eqnarray}
\label{multitrace_def}
  G\lr{n_1, \ldots, n_m} = \vev{ \frac{1}{N} \tr\phi^{2 n_1} \ldots \frac{1}{N} \tr\phi^{2 n_m} }
  =
  \sum\limits_{g=0}^{+\infty} \frac{1}{N^{2 g}}
  G_g\lr{n_1, \ldots, n_m} .
\end{eqnarray}
The Schwinger-Dyson equations can be derived in the conventional way by expanding the full derivative in the path integral, and take the following form:
\begin{eqnarray}
\label{phi4sd_nonplanar_Gg2}
 G_g\lr{2}    = \delta_{g,0} + \lambda G_g\lr{4}, \quad 
 G_g\lr{1, 1} = \delta_{g,1} + \lambda G_g\lr{3, 1} , \\
\label{phi4_sd_nonplanar_Gg1n}
 G_g\lr{1, n_2, \ldots, n_m} = \sum\limits_{k=2}^m \delta_{n_k, 1} G_{g-1}\lr{n_2, \ldots, n_{k-1}, n_{k+1}, \ldots n_m} + \lambda G_g\lr{3, n_2, \ldots, n_m}
 + \nonumber \\ +
 \sum\limits_{k=2}^{m} n_k \lr{1 - \delta_{n_k, 1}} G_{g-1}\lr{n_2, \ldots, n_{k-1}, n_k - 1, n_{k+1}, \ldots, n_m} , \\
\label{phi4sd_nonplanar_Ggn}
 G_g\lr{n_1, \ldots, n_m}
 =
 2 G_g\lr{n_1 - 2, n_2, \ldots, n_m}
 + \nonumber \\ +
 \sum\limits_{a=1}^{n_1-2}
 G_g\lr{a, n_1 - 2 - a, n_2, \ldots, n_m}
 +
 \lambda G_g\lr{n_1 + 2, n_2, \ldots, n_m}
 + \nonumber \\ +
 \sum\limits_{k=2}^{m} n_k
 G_{g-1}\lr{n_2, \ldots, n_{k-1}, n_k + n_1 - 2, n_{k+1}, \ldots, n_m}, \quad n_1 > 1 .
\end{eqnarray}
Correspondingly, the states of our DiagMC algorithm (variables $X$ in the notation of the previous Sections) can be described by sequences of positive integers $\lrc{n_1, \ldots, n_m}_g$ of any length $m$, labelled by the variable $g \geq 0$ (which has a geometric interpretation in terms of diagram genus). Note that for positive $\lambda$ all the terms on the right-hand side of equations (\ref{phi4sd_nonplanar_Gg2}), (\ref{phi4_sd_nonplanar_Gg1n}) and (\ref{phi4sd_nonplanar_Ggn}) are positive, thus no sign reweighting is necessary.

 A subtle point in the implementation of Diagrammatic Monte-Carlo for the Schwinger-Dyson equations (\ref{phi4sd_nonplanar_Gg2})-(\ref{phi4_sd_nonplanar_Gg1n}) is that the quantities $G_g\lr{n_1, \ldots, n_m}$ are not normalizable due to factorial growth of the number of contributing diagrams with the genus $g$ and hence cannot be directly interpreted as statistical weights. On the other hand, at fixed $g$ the number of contributing diagrams grows only exponentially with the number of vertices and external legs. For fixed $\lrc{n_1, \ldots, n_m}$ this growth is compensated by the powers of the coupling $\lambda$ if $\lambda$ is smaller than the critical value $\lambda_c = 1/12$. To deal with the growth of $G_g\lr{n_1, \ldots, n_m}$ at large $g$, $n$ and $m$ we assume that $G_g\lr{n_1, \ldots, n_m}$ are proportional to the probability $w_g\lr{n_1, \ldots, n_m}$ to find the state $\lrc{n_1, \ldots, n_m}_g$ in the Monte-Carlo process times the normalization factor $\lr{2 g}! \mathcal{\eta}^m {\kappa}^{n_1 + \ldots + n_m} $ which compensates for this growth. After such a redefinition of variables, we can directly identify the Metropolis updates and their weights from the Schwinger-Dyson equations (\ref{phi4sd_nonplanar_Gg2}), (\ref{phi4_sd_nonplanar_Gg1n}) and (\ref{phi4sd_nonplanar_Ggn}):
\begin{description}
  \item[Create single trace:] $\lrc{n_1, \ldots, n_m}_g \rightarrow \lrc{2, n_1, n_2, \ldots, n_m}_g$, weight $A\lr{X | Y} = \lr{\kappa^2 \eta}^{-1}$.
  \item[Create two traces:] $\lrc{n_1, \ldots, n_m}_g \rightarrow \lrc{1, n_1, n_2, \ldots, n_{k-1}, 1, n_{k}, \ldots, n_m}_{g+1}$, $k$ takes random value in the range $2 \ldots m+1$, weight $A\lr{X | Y} = \frac{\lr{m+1}}{\lr{2 g + 2}\lr{2 g + 1}} \frac{1}{\eta^2 \kappa^2}$. The genus $g$ is increased by one. To ``undo'' this update, one should store in memory the position $k$ at which the second trace was inserted.
  \item[Insert line:] $\lrc{n_1, \ldots, n_m}_g \rightarrow \lrc{n_1 + 2, n_2, \ldots, n_m}_g$, weight $A\lr{X | Y} = 2/\kappa^2$.
  \item[Merge two traces:] $\lrc{n_1, \ldots, n_m}_g \rightarrow \lrc{n_1 + n_2 + 2, n_3, \ldots, n_m}_g$, weight $A\lr{X | Y} = \eta/\kappa^2$. To ``undo'' this update, one should store in memory either $n_1$ or $n_2$.
  \item[Create vertex:] $\lrc{n_1, \ldots, n_m}_g \rightarrow \lrc{n_1 - 2, n_2, \ldots, n_m}_g$, weight $A\lr{X | Y} = \lambda \kappa^2$.
  \item[Split single trace:] $\lrc{n_1, \ldots, n_m}_g \rightarrow \lrc{a, n_1, \ldots, n_{k-1}, n_k + 2 - a, n_{k+1}, \ldots, n_m}_{g+1}$, $k$ takes random value in the range $1 \ldots m$, $a$ takes random value in the range $1 \ldots n_k + 1$, the genus $g$ is increased by one, weight $A\lr{X | Y} = \frac{\sum\limits_{k=1}^m \lr{n_k + 1}\lr{n_k + 2}}{\lr{2 g + 2}\lr{2 g + 1} \eta \kappa^2}$. To ``undo'' this update, one should store in memory the position $k$ of the trace which is being split.
\end{description}
The initial states of the algorithm are either $\lrc{2}_0$ or $\lrc{1,1}_1$ with probabilities being proportional to $\lr{\eta \kappa}^{-1}$ and $\lr{2 \eta^2 \kappa^2}^{-1}$, corresponding to the first terms in the right-hand sides of equations (\ref{phi4sd_nonplanar_Gg2}). These Metropolis updates provide a stochastic implementation of the so-called topological recursion \cite{Eynard:14:1} for the matrix model (\ref{phi4_matrix_model}).

\begin{figure*}
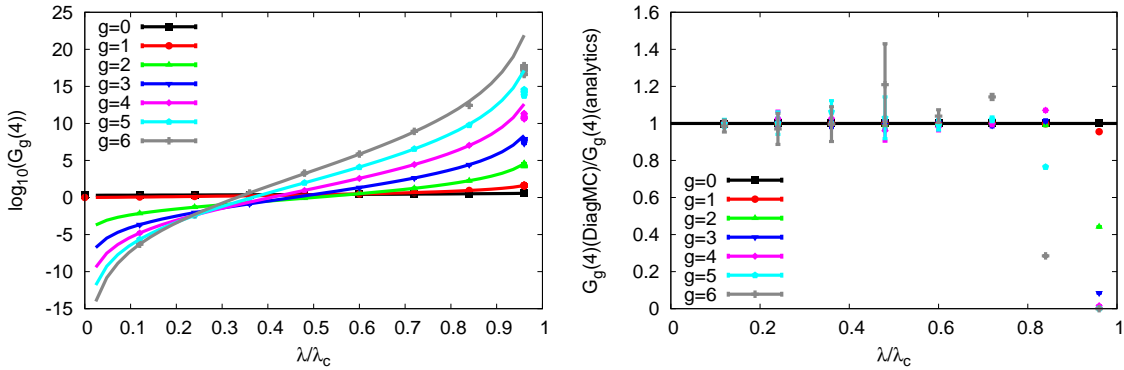

  \centering
  \includegraphics[angle=-90,width=0.5\textwidth]{{{G4}}}\includegraphics[angle=-90,width=0.5\textwidth]{{{G4_ratios}}}\\
  \caption{Comparison of the coefficients of the $1/N^2$ expansion of the expectation value $G_4 = \vev{\frac{1}{N} \tr \phi^4}$ in the $\phi^4$ matrix model (4.1) obtained from the DiagMC simulation (data points with error bars) with exact analytic results \cite{Marino:07:1} (solid lines).}
  \label{fig:G4}
\end{figure*}

 In order to test the performance of our Diagrammatic Monte-Carlo algorithm, we consider the expectation values $G_g\lr{4}$ which can be directly related to the $1/N$ expansion of the free energy in the matrix model (\ref{phi4_matrix_model}):
\begin{eqnarray}
\label{genus_expansion}
  - \frac{4}{N^2} \frac{\partial}{\partial \lambda} \log \mathcal{Z} = \vev{\frac{1}{N} \tr\phi^4} = \sum\limits_{g=0}^{+\infty} \frac{1}{N^{2 g}} G_g\lr{4} .
\end{eqnarray}
On Fig.~\ref{fig:G4} we compare the results for the coefficients $G_g\lr{4}$ obtained in the above described DiagMC algorithm with the analytic results obtained in \cite{Marino:07:1}. Each data point was obtained by averaging over $2 \cdot 10^9$ Metropolis updates, which took several hours on a single CPU core. While the absolute value of $G_g\lr{4}$ spans almost 30 orders of magnitude for $\lambda \in \lrs{0 \ldots \lambda_c}$ and $g = 0 \ldots 6$ (see left plot on Fig.~\ref{fig:G4}), the ratios between the outcome of Monte-Carlo sampling and analytic results are equal to one within statistical errors (see right plot on Fig.~\ref{fig:G4}), except for the data points at higher genera $g$ and the values of $\lambda$ close to $\lambda_c$, for which the Monte-Carlo data is significantly smaller than the exact result. Analysis of the histograms of $G_g\lr{4}$ for these data points suggests that the origin of under-sampling lies in the heavy-tailed distribution of observables, for which the expectation value should be saturated by rare extreme values. The required increase of statistics reflects the critical slowing down of the algorithm near the critical value $\lambda = \lambda_c$, where the geometric series (\ref{sd_linear_powerseries_explicit}) are at the edge of convergence.


\end{document}